\documentclass[12pt,epsbox]{article}
\usepackage[dvips]{graphicx}

\makeatletter
\@addtoreset{equation}{section}
\renewcommand{\theequation}{\thesection.\arabic{equation}}

\newcommand{\N}{{\bf N}}

\newcommand{\bR}{{\bf R}}

\newcommand{\bS}{{\bf S}}
\newcommand{\bZ}{{\bf Z}}

\newcommand{\cN}{{\cal N}}

\newcommand{\nn}{\nonumber \\}

\newcommand{\tr}{{\rm tr}\,}

\newcommand{\be}{\begin{equation}} \newcommand{\ee}{\end{equation}}
\newcommand{\bea}{\begin{eqnarray}} \newcommand{\eea}{\end{eqnarray}}

\font\zfont = cmss10 
\newcommand{\ZZ}{\hbox{\zfont Z\kern-.4emZ}}

\ifx\TwoupWrites\UnDeFiNeD\else\target{\magstepminus1}{11.3in}{8.27in}
\source{\magstep0}{7.5in}{11.69in}\fi
\newfont{\fourteencp}{cmcsc10 scaled\magstep2}
\newfont{\titlefont}{cmbx10 scaled\magstep3}
\newfont{\authorfont}{cmcsc10 scaled\magstep1}
\newfont{\fourteenmib}{cmmib10 scaled\magstep2}
\skewchar\fourteenmib='177
\newfont{\elevenmib}{cmmib10 scaled\magstephalf}
\skewchar\elevenmib='177
\makeatletter
\newcommand\nonsequentialeqnum{
\@addtoreset{equation}{section}
\def\theequation{\arabic{section}.\arabic{equation}}}
\newif\ifp@bblock \p@bblocktrue
\newcommand\nopubblock{\p@bblockfalse}
\newcommand\topspace{\hrule height 0pt depth 0pt \vskip}
\newcommand\p@bblock{\begingroup \tabskip=\hsize minus \hsize
\baselineskip=1.5\ht\strutbox \topspace-2\baselineskip
\halign to\hsize{\strut ##\hfil\tabskip=0pt\crcr
\the\Pubnum\crcr\the\date\crcr}\endgroup}

\renewcommand\titlepage{\ifx\TwoupWrites\UnDeFiNeD\null
\vspace{-1.7cm}\fi
\vskip0.6cm
\ifp@bblock\p@bblock \else\hrule height 0pt \relax \fi}
\makeatother
\newtoks\date
\newtoks\Pubnum
\newtoks\pubnum
\newcommand{\frontpageskip}{\vspace{12pt plus .5fil minus 2pt}}
\renewcommand{\title}[1]{\frontpageskip
\begin{center}{\titlefont #1}\end{center}\par}
\renewcommand{\author}[1]{\frontpageskip\par\begin{center}
{\authorfont #1}\end{center}
\nobreak
}

\renewcommand{\thanks}[1]{\footnote{#1}}
\renewcommand{\abstract}{\par\frontpageskip\centerline{
\fourteencp Abstract}
\vspace{8pt plus 3pt minus 3pt}}

\begin{document}

\begin{titlepage}
\hfill
\vbox{
    \halign{#\hfil         \cr
           CERN-TH/2002-048 \cr
           TAUP-2699-02 \cr
           hep-th/0203052  \cr
           } 
      }  
\vspace*{20mm}
\begin{center}
{\Large {\bf  Comments on M Theory Dynamics\\ 
on $G_2$ Holonomy Manifolds}\\} 
\vspace*{15mm}
{\sc Harald Ita,}$^{a}$ 
\footnote{e-mail: {\tt ita@post.tau.ac.il}}
{\sc Yaron Oz}$^{a\,b}$ 
\footnote{e-mail: {\tt yaronoz@post.tau.ac.il, Yaron.Oz@cern.ch}}
and {\sc Tadakatsu Sakai}$^{a}$
\footnote{e-mail: {\tt tsakai@post.tau.ac.il}}

\vspace*{1cm} 
{\it {$^{a}$ Raymond and Beverly Sackler Faculty of Exact Sciences\\
School of Physics and Astronomy\\
Tel-Aviv University , Ramat-Aviv 69978, Israel}}\\ 

\vspace*{5mm}
{\it {$^{b}$Theory Division, CERN \\
CH-1211 Geneva  23, Switzerland}}\\

\end{center}

\begin{abstract}

We study the dynamics of M-theory on $G_2$ holonomy manifolds,
and consider in detail the manifolds
 realized as the quotient of 
the spin bundle over $\bS^3$ by discrete groups.
We analyse, in particular, the class of quotients 
where the triality symmetry is
broken. 
We study the structure of the moduli space, 
construct its defining equations and show that 
three different types of classical geometries are interpolated 
smoothly. 
We derive the $\cN=1$ superpotentials of M-theory on 
the quotients and 
comment on the
membrane instanton physics. 
Finally, we turn on Wilson lines that break 
gauge symmetry and  
discuss some of the implications.

\end{abstract}
\vskip 2cm

March 2002

\end{titlepage}

\newpage

\section{Introduction}

The study of the dynamics of M-theory on $G_2$ holonomy
manifolds has been receiving much attention recently. 
It provides  
a framework for the analysis of ${\cal N}=1$ supersymmetric systems 
in four dimensions. 
When the $G_2$ holonomy manifolds are smooth and large
the low energy physics can be described by a
Kaluza-Klein reduction of eleven-dimensional supergravity 
(see {\it e.g.} \cite{paptow}). 
These cases are the less interesting ones 
since at low energy we find neither 
non-abelian gauge symmetry nor chiral matter.
For these we have to consider  
singular $G_2$ manifolds.
 
Recently, there has been a significant
 progress in understanding the dynamics of
 M-theory on singular $G_2$ manifolds \cite{ach0}-\cite{achvaf}. 
For instance, appropriate singular quotients of 
$S(\bS^3)$, the spin bundle over $\bS^3$, 
yield $\cN=1$ super Yang-Mills (SYM) theory 
with an $ADE$ gauge group \cite{bs}\cite{gpp}. 
It has been  shown that M-theory on these backgrounds reproduces 
the vacuum structure of the field theory. 
It was demonstrated in \cite{amv}, when studying the 
duality proposed in \cite{vafa},
that a singular quotient of $S(\bS^3)$ can 
be interpolated smoothly to a non-singular quotient of $S(\bS^3)$ 
using a $G_2$ flop. 
For other work on various aspects of $G_2$ holonomy compactifications
and branes, 
see \cite{Acharya:2000mu}-\cite{cft}. 

The dynamics of M-theory on $S(\bS^3)$ and its quotients was
discussed 
in detail in \cite{atiwit}. 
It was shown that 
the moduli space of the theory is given by 
a complex one-dimensional manifold that interpolates between three different 
$S(\bS^3)$ smoothly. 
They also confirmed the result of \cite{amv} by showing that 
the moduli space of M-theory  
interpolates smoothly between a  singular quotient of $S(\bS^3)$ 
that yields pure SYM at low energy and
 a non-singular quotient of $S(\bS^3)$. 

The purpose of this paper is to extend the results of \cite{atiwit}
to more general quotients of $S(\bS^3)$. 
We will mainly analyse a ``double'' quotient, 
where the quotient group is a subgroup of the isometry of 
$S(\bS^3)$. We will
study the moduli space of M-theory on the quotient and show
that the moduli space interpolates smoothly between three different 
types of classical geometries.
These three geometries yield different low energy dynamics. 
Thus, the result can be viewed as another example of a $G_2$ flop. 
We will also compute the exact superpotentials of M-theory on the
quotients of 
$S(\bS^3)$, and  
comment on the
membrane instanton physics. 
An interesting feature of the models that we study is the possibility
of turning on 
Wilson lines that break gauge symmetry (see also \cite{wit:wl}). 
We will discuss some of the implications.

The paper is organized as follows.
In section 2 we review the geometry of $S(\bS^3)$ 
and its quotients. 
In section 3 we will construct the complex curve
that describes the moduli space 
of M-theory on a double quotient. 
We will show that the curve interpolates smoothly between
three different classical 
geometries, and suggest a new 
$G_2$ flop. 
In section 4 we compute exact superpotentials of M-theory 
on quotients of $S(\bS^3)$, 
and discuss some implications to the membrane instanton 
physics. 
In section 5 we will make some comments on
 M-theory on the double quotient with 
Wilson lines.
Section 6 is devoted to a discussion
of open problems.

\section{The spin bundle $S(\bS^3)$} 

The seven-dimensional spin bundle $S(\bS^3)$
is asymptotic at infinity to a cone over the six-dimensional
space 
$Y=\bS^3\times \bS^3=SU(2)^3/SU(2)$. 
The Einstein metric on $Y$ 
can be constructed as follows \cite{atiwit}. 
Let 
$(g_1,g_2,g_3)\in SU(2)^3$ be a group element, and 
impose the equivalence relation 
\begin{equation}
(g_1,g_2,g_3)\cong (g_1h,g_2h,g_3h),~~h\in SU(2) \ . 
\label{equiv}
\end{equation}
Define
$a=g_2g_3^{-1},b=g_3g_1^{-1},c=g_1g_2^{-1}$  
($abc=1$). The Einstein metric on $Y$ is given by 
\begin{equation}
d\Omega_Y^2=da^2+db^2+dc^2 \ , 
\end{equation}
where $da^2=-{\rm tr}\,(a^{-1}da)^2$ etc.
 
The metric on $Y$ admits the isometry group 
$SU(2)_u\times SU(2)_v\times SU(2)_w$: 
\begin{equation}
g_1\rightarrow ug_1,~~g_2\rightarrow vg_2,~~
g_3\rightarrow wg_3 \ , 
\end{equation}
or equivalently 
\begin{equation}
a\rightarrow vaw^{-1},~~b\rightarrow wbu^{-1},~~
c\rightarrow ucv^{-1} \ . 
\end{equation}
In addition, the metric admits the triality symmetry $\Sigma_3$ 
generated by 
\begin{eqnarray} 
&&\alpha : (a,b,c)\rightarrow (c^{-1},b^{-1},a^{-1})\ , \nonumber\\
&&\beta : (a,b,c)\rightarrow (b,c,a) \ . 
\end{eqnarray}
Note that $\Sigma_3$ acts on $(g_1,g_2,g_3)$ by 
permutations. 

We can extend the cone on $Y$ to the interior 
and obtain a seven-dimensional $G_2$ holonomy manifold. 
In fact one obtains three different 
$G_2$ holonomy manifolds which differ
by the $\bZ_2 \subset\Sigma_3$ symmetry group.
They are:
\begin{eqnarray}
&&
X_1,X_1^{\prime}:~~~~~
ds^2={dr^2\over 1-(r_0/r)^3}+
{r^2\over 72}\left(1-(r_0/r)^3\right)(2dc^2-da^2+2db^2)+
{r^2\over 24}da^2 \ , \nonumber\\ 
&&X_2,X_2^{\prime}:~~~~~
ds^2={dr^2\over 1-(r_0/r)^3}+
{r^2\over 72}\left(1-(r_0/r)^3\right)(2da^2-db^2+2dc^2)+
{r^2\over 24}db^2 \ , \nonumber\\ 
&&X_3,X_3^{\prime}:~~~~~
ds^2={dr^2\over 1-(r_0/r)^3}+
{r^2\over 72}\left(1-(r_0/r)^3\right)(2db^2-dc^2+2da^2)+
{r^2\over 24}dc^2 \ . 
\end{eqnarray}
Here $r_0$ is a constant which corresponds
to the size of the base $\bS^3$. 

The $\bZ_2$ isometry groups of $X_i$, $\bZ_2^{(i)}$, act as 
\begin{eqnarray}
\bZ_2^{(1)}:~(a,b,c)\rightarrow (a^{-1},c^{-1},b^{-1}) \ , \nonumber\\
\bZ_2^{(2)}:~(a,b,c)\rightarrow (c^{-1},b^{-1},a^{-1}) \ , \nonumber\\
\bZ_2^{(3)}:~(a,b,c)\rightarrow (b^{-1},a^{-1},c^{-1}) \ . 
\end{eqnarray}
Note that $SU(2)_u\times SU(2)_v\times SU(2)_w$ is also 
the isometry group of $X_i$.

It is useful to rewrite these metrics in the form of 
\cite{gpp} 
\begin{equation}
ds^2={dr^2\over 1-(r_0/r)^3}+
{r^2\over 9}\left(1-(r_0/r)^3\right)\left(\sigma^a-{1\over 2}\Sigma^a\right)^2
+{r^2\over 12}\left(\Sigma^a\right)^2 \ . 
\label{metric;spinb}
\end{equation}
Here $\sigma^a,~a=1,2,3,$ are the left-invariant one-forms on $\bS^3$ 
in an $\bR^4$ fibration, and $\Sigma^a$ the
left-invariant one-forms
on the base $\bS^3$. 

It admits the isometry
\begin{equation}
SU(2)_L\times \widetilde{SU(2)}_L \times SU(2)_R^{diag} \ ,
\label{isometry;amv}
\end{equation}
where $SU(2)_L\times SU(2)_R$ and 
$\widetilde{SU(2)}_L\times \widetilde{SU(2)}_R$ are the isometry 
groups for the fiber $\bS^3$ and the base $\bS^3$, respectively. 
Here $SU(2)_R^{diag}$ is the diagonal subgroup of 
$SU(2)_R\times \widetilde{SU(2)}_R$.
The associative three-form of this $G_2$ holonomy
manifold is given by \cite{amv}
\begin{equation}
\Omega={r_0^3\over 72}\,\epsilon_{abc}\,\Sigma^a\Sigma^b\Sigma^c
+d\left({r^3-r_0^3\over 9}\,\Sigma^a\sigma^a\right) \ . 
\label{omega}
\end{equation}
%
%
This is invariant
under the isometry group (\ref{isometry;amv}).
Therefore, any quotient of (\ref{metric;spinb}) 
by a finite subgroup 
of the isometry group  (\ref{isometry;amv}) 
preserves the $G_2$ structure. 

We write the left-invariant one-forms in the form 
\begin{equation} 
\sigma^a=-{i\over 2}\,\tr (T^a g^{-1}dg),~~~~~~~
\Sigma^a=-{i\over 2}\,\tr (T^a\tilde{g}^{-1} d\tilde{g}) \ , 
\end{equation}
with $g,\tilde{g}$ being the $SU(2)$ group elements of the fiber and 
base $\bS^3$, respectively. 
$T^a$ are $SU(2)$ generators with the normalization 
$\tr (T^aT^b)=2\delta^{ab}$.
We can label $X_i,X_i^{\prime}$ as

\begin{itemize}
\item $X_1$: 
\begin{equation}
g=b^{-1},~~\tilde{g}=a, 
\label{x1:g}
\end{equation}
\begin{equation}
SU(2)_L=SU(2)_u,~~ \widetilde{SU(2)}_L=SU(2)_v,~~
SU(2)_R^{diag}=SU(2)_w \ . 
\end{equation}

\item $X_1^{\prime}$: 
\begin{equation}
g=c,~~\tilde{g}=a^{-1}, 
\end{equation}
\begin{equation}
SU(2)_L=SU(2)_u,~~ \widetilde{SU(2)}_L=SU(2)_w,~~
SU(2)_R^{diag}=SU(2)_v \ . 
\end{equation}

\item $X_2$: 
\begin{equation}
g=a,~~\tilde{g}=b^{-1}, 
\end{equation}
\begin{equation}
SU(2)_L=SU(2)_v,~~ \widetilde{SU(2)}_L=SU(2)_u,~~
SU(2)_R^{diag}=SU(2)_w \ . 
\end{equation}

\item $X_2^{\prime}$: 
\begin{equation}
g=c^{-1},~~\tilde{g}=b, 
\end{equation}
\begin{equation}
SU(2)_L=SU(2)_v,~~ \widetilde{SU(2)}_L=SU(2)_w,~~
SU(2)_R^{diag}=SU(2)_u \ . 
\end{equation}

\item $X_3$: 
\begin{equation}
g=a^{-1},~~\tilde{g}=c, 
\end{equation}
\begin{equation}
SU(2)_L=SU(2)_w,~~ \widetilde{SU(2)}_L=SU(2)_u,~~
SU(2)_R^{diag}=SU(2)_v \ .
\end{equation}

\item $X_3^{\prime}$: 
\begin{equation}
g=b,~~\tilde{g}=c^{-1}, 
\end{equation}
\begin{equation}
SU(2)_L=SU(2)_w,~~ \widetilde{SU(2)}_L=SU(2)_v,~~
SU(2)_R^{diag}=SU(2)_u \ . 
\end{equation}
\end{itemize}

Consider next the homology three-cycles of these $G_2$ 
holonomy manifolds. 
Following \cite{atiwit}, let $\hat{D}_i\subset SU(2)^3$ 
be the $i^{\rm th}$ copy of $SU(2)$ given by 
$(g_i,g_{i+1},g_{i+2})=(g,1,1)$, 
where the index $i$ is defined mod $3$. 
In $Y=SU(2)^3/SU(2)$, the $\hat{D}_i$ project to three-cycles 
$D_i$. 
Since $b_2(Y)=2$, $D_i$ are not independent and obey the relation 
\begin{equation}
D_1+D_2+D_3=0 \ .
\end{equation}
The triality $\Sigma_3$ acts on 
the $D_i$ by permutations. 
In terms of $a,b,c$, the $D_i$ are given by 
\begin{eqnarray}
&&D_1:~a=1=bc, \nn
&&D_2:~b=1=ca, \nn
&&D_3:~c=1=ab \ . 
\end{eqnarray}
It was  argued in \cite{atiwit} that one obtains different $G_2$ holonomy 
manifolds depending on which three-cycle is ``filled in''. 
As an example, consider $X_1$
which is defined by filling in $D_1$,  
that is, $D_1$ is identified with the fiber $\bS^3$.  
Indeed, it follows from $a=1=bc$ and (\ref{x1:g}) that 
$\sigma^a\ne 0$ while $\Sigma^a$ vanish. 
The base $\bS^3$ should correspond to the 
three-cycle of $b=1$, namely $D_2$ 
because $\sigma^a=0,~\Sigma^a\ne 0$. 
In general we get
\begin{equation}
\begin{array}{llll}
X_1:&(b^{-1},a;D_2) & X_1^{\prime}:&(c,a^{-1};-D_3) \\
X_2:&(a,b^{-1};-D_1) &X_2^{\prime}:&(c^{-1},b;D_3)  \\
X_3:&(a^{-1},c;D_1) & X_3^{\prime}:&(b,c^{-1};-D_2) \ , 
\end{array}
\end{equation}
where we label
$g$,$\tilde{g}$ 
and $Q$ the associative three-cycle (base).
Recall that each $\bZ_2$ acts by a parity transformation of $X_i$. 

\subsection{Quotients of  $S(\bS^3)$} 

We have seen that any quotient of these $G_2$ holonomy  manifolds by 
a finite subgroup of the isometry preserves the $G_2$ structure. 
However, the triality $\Sigma_3$ of the base $Y$ 
is broken by the quotient.
As an example, consider a quotient by $\Gamma_1\subset SU(2)_u$.
We impose the  equivalence relation 
\begin{equation}
b\sim b\gamma_1^{-1},~~c\sim\gamma_1c \ , 
\end{equation}
with $\gamma_1\in\Gamma_1$. 
We will refer to this as a single quotient. 
The quotient leaves unbroken only the subgroup $\bZ_2^{(1)}$. 
As noted in \cite{atiwit}, this symmetry plays an 
important role in determining  the curve describing the moduli space. 
We will mainly discuss the double
quotient by 
\begin{equation}
\Gamma_1\subset SU(2)_u,~~\Gamma_2\subset SU(2)_v \ . 
\end{equation}
The equivalence relation to be imposed is 
\begin{equation}
a\sim\gamma_2 a,~~b\sim b\gamma_1^{-1},~~c\sim\gamma_1c\gamma_2^{-1} \ . 
\label{equiv;double}
\end{equation}
It follows that $Y_{\Gamma_1\times\Gamma_2}$, the quotient of 
$Y$, preserves none of the triality symmetry. 

Let us consider the double quotient of $S(\bS^3)$ in detail. 
We first discuss the homology of the base 
$Y_{\Gamma_1\times\Gamma_2}$. 
To this end, we start from the three-cycles $D_i$ in $Y$ and 
see how the quotient acts on them. 
$D_3$ projects to a cycle $D^{\prime}_3=\bS^3$. 
On the other hand, $D_i,~i=1,2$ project to 
$N_i$-fold covers of $D_i^{\prime}=\bS^3/\Gamma_i$
with $N_i$ the order of the finite group $\Gamma_i$. 
These cycles obey the relation
\begin{equation}
N_1D_1^{\prime}+N_2D_2^{\prime}+D_3^{\prime}=0 \ . 
\label{cycle:double}
\end{equation}

Let us now discuss the topological structure of the double
quotient of $S(\bS^3)$. 
For instance, the quotient of $X_1: (b^{-1},a;D_2)$ 
is topologically 
\begin{equation}
X_{1,\Gamma_1\times\Gamma_2}=\bR^4/\Gamma_1\times \bS^3/\Gamma_2 \ . 
\end{equation}
It follows from (\ref{equiv;double}) that 
the base manifold is homologous to $D_2^{\prime}$. 
The topologies of the quotients are summarized by:
\begin{equation}
\begin{array}{ll}
X_{1,\Gamma_1\times\Gamma_2}=\bR^4/\Gamma_1\times \bS^3/\Gamma_2: & 
(b^{-1},a;{D_2^{\prime}}) \\ 
X_{1,\Gamma_1\times\Gamma_2}^{\prime}=\bR^4/\Gamma_1\times \bS^3/\Gamma_2: & 
(c,a^{-1};-{D_3^{\prime}\over N_2}) \\
X_{2,\Gamma_1\times\Gamma_2}=\bR^4/\Gamma_2\times \bS^3/\Gamma_1: & 
(a,b^{-1};-D_1^{\prime}) \\ 
X_{2,\Gamma_1\times\Gamma_2}^{\prime}=\bR^4/\Gamma_2\times \bS^3/\Gamma_1:&
(c^{-1},b;{D_3^{\prime}\over N_1})  \\
X_{3,\Gamma_1\times\Gamma_2}=\bR^4\times \bS^3/(\Gamma_1\times\Gamma_2):& 
(a^{-1},c;{D_1^{\prime}\over N_2}) \\ 
X_{3,\Gamma_1\times\Gamma_2}^{\prime}=
\bR^4\times \bS^3/(\Gamma_1\times\Gamma_2):&
(b,c^{-1};-{D_2^{\prime}\over N_1}) \ .
\label{topology:d}
\end{array}
\end{equation}
Recall that 
$X_{i,\Gamma_1\times\Gamma_2}$ and 
$X_{i,\Gamma_1\times\Gamma_2}^{\prime}$ 
can be obtained by filling in $D_i^{\prime}$ of 
$Y_{\Gamma_1\times\Gamma_2}$. 
For instance, $X_{1,\Gamma_1\times\Gamma_2}$ has the associative
three-cycle $Q=D_2^{\prime}$. 
One finds from (\ref{cycle:double}) that the $Q$ 
is homologous to $-D_3^{\prime}/N_2$, which is the associative 
three-cycle of $X_{1,\Gamma_1\times\Gamma_2}^{\prime}$.

\section{M-theory on double quotients}

In this section we construct the curve
describing the moduli space of M-theory
on the double quotients
defined by
(\ref{equiv;double}).

Let ${\cal N}_{\Gamma_1\times\Gamma_2}$ be 
the space of parameters of M-theory on the double quotients. 
Denote by 
 $P_i\in {\cal N}_{\Gamma_1\times\Gamma_2}$ the points 
that correspond to the large volume limit of 
$X_{i,\Gamma_1\times\Gamma_2},X_{i,\Gamma_1\times\Gamma_2}^{\prime}$.
The local coordinates of ${\cal N}_{\Gamma_1\times\Gamma_2}$ near 
these points are 
\begin{eqnarray}
&&\eta_1=\exp\left(  {2k\over 3N_1}f_3+{k\over 3N_1}f_1
+i\alpha_1^{\prime}\right), \nn
&&\eta_2=\exp\left(  {2k\over 3N_2}f_1+{k\over 3N_2}f_2
+i\alpha_2^{\prime}\right), \nn
&&\eta_3=\exp\left(  {2k\over 3}f_2+{k\over 3}f_3
+i\alpha_3^{\prime}\right) \ , 
\label{def:eta}
\end{eqnarray}
with 
\begin{equation}
\alpha_i^{\prime}=\int_{D_i^{\prime}}C \ .
\end{equation}
Here $k$ is a constant that can be determined using an instanton 
correction. 

At $P_i$, we have 
\begin{eqnarray}
(f_i,f_{i+1},f_{i-1})=\rho\,(-2,1,1) \ ,~~~\rho\rightarrow\infty \ ,
\end{eqnarray}
and thus 
\begin{equation}
\eta_i=e^{i\alpha_i^{\prime}},~~\eta_{i+1}=0,~~\eta_{i-1}=\infty \ . 
\end{equation}

In order to determine the orders of the pole and zero we
will use the
membrane instanton corrections(for a discussion on membrane instantons, see
{\it e.g.} \cite{hm}). 
One-instanton action is given by 
\begin{equation}
u=\exp\left( -V(Q)+i\int_{Q}C\right) \ , 
\end{equation}
where $Q$ is an associative three-cycle and $V(Q)$ is its volume.

Consider the different points:
\begin{itemize}
\item $P_1$:  
$X_{1,\Gamma_1\times\Gamma_2}$. 
A non-trivial $C$-field that breaks the
gauge symmetry can be turned on 
along an ALE fibration 
by applying fiberwise the duality between M-theory on K$3$ 
and heterotic string on $T^3$. 
In this case, we have 
\begin{equation}
\int_{D_1^{\prime}}C={2\pi\mu_1\over t_1} \ . 
\end{equation}
Note that this corresponds to a triple in the heterotic string dual 
\cite{ddh}. 
Thus we find 
\begin{equation}
\eta_1=e^{2\pi i\mu_1/t_1} \ . 
\end{equation}

By comparing $\eta_2,\eta_3$ to $u$ with 
$Q=D_2^{\prime}=-{1\over N_2}D_3^{\prime}$, 
we find 
\begin{equation}
\eta_2\sim u^{+1},~~\eta_3\sim u^{-N_2} \ ,
\end{equation}
where we used $D_1^{\prime}=0$, which holds since 
this cycle is filled in. 
We now recall the ansatz of \cite{atiwit} that 
an M2-brane wrapping an associative three-cycle with 
$\int C=2\pi\mu/t$ is a SYM instanton of $K_t\subset G_{\Gamma}$ 
with instanton number $t$. 
Here $G_{\Gamma}$ is an $ADE$ group 
corresponding to the finite group $\Gamma$. 
$t$ is a positive integer that divides some of the Dynkin indices 
$k_i$ of $G_{\Gamma}$. 
$K_t$ is a subgroup 
of $G_{\Gamma}$ that is left unbroken by the $C$-field, and 
defined by Dynkin indices whose elements consist of a set of 
integer $k_i/t$. 
Then a good local coordinate around $P_1$ should be taken to be 
$u^{1/t_1h_{t_1}}$ which is equal to the gaugino bilinear 
condensate of SYM with the gauge group $K_{t_1}$. 
Here $h_{t_1}$ is the dual Coxeter number of $K_{t_1}$. 
Therefore, we find that $\eta_2$ at $P_1$ has zeros of order 
$t_1h_{t_1}$ and $\eta_3$ has poles of order $N_2t_1h_{t_1}$. 


\item $P_2$: 
As in the case of $P_1$, one can turn on a $C$-field on
$D_2^{\prime}$ so that 
\begin{equation}
\eta_2=e^{2\pi i\mu_2/t_2} \ . 
\end{equation}
$\eta_3$ has zeros of order $N_1t_2h_{t_2}$, and $\eta_1$ has poles of 
order $t_2h_{t_2}$ since 
$Q=-D_1^{\prime}={1\over N_1}D_3^{\prime}$.

\item $P_3$: 
\begin{equation}
\eta_3=1,~~\eta_1\propto \exp\left( i\int_{D_1^{\prime}}C\right),~~
\eta_2\propto \exp\left(i\int_{D_2^{\prime}}C\right) \ . 
\end{equation}
Since the associative three-cycle is given by 
$Q=D_1^{\prime}/N_2=-D_2^{\prime}/N_1$, 
we find that 
\begin{equation}
\eta_1=0^{N_2},~~\eta_2=\infty^{N_1} \ . 
\end{equation}

\end{itemize}
To summarize, the behaviour of $\eta_i$ at the points $P_i$ is
given by:
\begin{equation}
\begin{array}{|c|c|c|c|} \hline 
 & P_1^{t_1,\mu_1}  & 
   P_2^{t_2,\mu_2}  & P_3 \\ \hline
\eta_1 & e^{2\pi i\mu_1/t_1} & 
 \infty^{t_2h_{t_2}} & 
0^{N_2} \\ \hline
\eta_2 & 0^{t_1h_{t_1}} & 
 e^{2\pi i\mu_2/t_2} & \infty^{N_1} \\ \hline
\eta_3 & \infty^{N_2t_1h_{t_1}} & 
 0^{N_1t_2h_{t_2}} & 1 \\ \hline
\end{array}
\label{table;double}
\end{equation}

\subsection{The M-theory curve}

Now we are ready to construct the curve describing the moduli space. 
One finds from the table (\ref{table;double}) that $\eta_i$ have 
the same order of zeros and poles. Thus,  $\eta_i$ can be regarded 
as meromorphic functions on a sphere. 
Let $z$ be the coordinate of this sphere. 
$\eta_i$ take the form 
\begin{eqnarray}
&&\eta_1=c_1\,{(z-\gamma)^{N_2}\over 
             \prod_{t_2,\mu_2}(z-\beta_{t_2,\mu_2})^{t_2h_{t_2}}} \ , \\
&&\eta_2=c_2\,{\prod_{t_1,\mu_{1}}(z-\alpha_{t_1,\mu_1})^{t_1h_{t_1}}
             \over (z-\gamma)^{N_1}} \ , \\
&&\eta_3=c_3\,{\prod_{t_2,\mu_2}(z-\beta_{t_2,\mu_2})^{N_1t_2h_{t_2}}\over 
               \prod_{t_1,\mu_1}(z-\alpha_{t_1,\mu_1})^{N_2t_1h_{t_1}}} \ , 
\end{eqnarray}
where $P_1^{t_1,\mu_1},P_2^{t_2,\mu_2},P_3$ are mapped to the points 
$z=\alpha_{t_1,\mu_1},\beta_{t_2,\mu_2},\gamma$, respectively. 
In addition we have to impose the relations 
\begin{eqnarray}
e^{2\pi i\mu_1/t_1}&\!=\!&c_1\,{(\alpha_{t_1,\mu_1}-\gamma)^{N_2}\over 
 \prod_{t_2,\mu_2}(\alpha_{t_1,\mu_1}-\beta_{t_2,\mu_2})^{t_2h_{t_2}}} \ , \nn
e^{2\pi i\mu_2/t_2}&\!=\!&c_2\,{
\prod_{t_1,\mu_1}(\beta_{t_2.\mu_2}-\alpha_{t_1,\mu_1})^{t_1h_{t_1}}\over 
(\beta_{t_2,\mu_2}-\gamma)^{N_1}} \ , \nn
1&\!=\!&c_3\,{\prod_{t_2,\mu_2}(\gamma-\beta_{t_2.\mu_2})^{N_1t_2h_{t_2}}
\over \prod_{t_1,\mu_1}(\gamma-\alpha_{t_1,\mu_{t_1}})^{N_2t_1h_{t_1}}} \ . 
\label{parameters}
\end{eqnarray}
From these one can see that 
\begin{equation}
c_1^{N_1}c_2^{N_2}c_3=(-1)^{N_1N_2} \ , 
\end{equation}
where we used the formula $\sum_{t,\mu}th_t=N$, with $N$ being the 
order of the finite group. 
It thus follows that 
\begin{equation}
\eta_1^{N_1}\eta_2^{N_2}\eta_3=(-1)^{N_1N_2} \ . 
\end{equation}

From this and (\ref{def:eta}), one finds that 
\begin{equation}
N_1\alpha_1^{\prime}+N_2\alpha_2^{\prime}+\alpha_3^{\prime}
=N_1N_2\pi ~~{\rm mod}~~2\pi. 
\label{anomaly}
\end{equation}
As in \cite{atiwit}, we should be able
to reproduce this relation using 
an anomaly argument. 
We will leave this as an open problem.
 
The existence of the curve shows that the moduli space of the 
double quotients interpolates smoothly between the three different classical 
geometries $P_1,P_2,P_3$. 
We can express the curve in terms of $\eta_i$ 
by eliminating $z$. 
For simplicity, consider the case 
$\Gamma_1=\bZ_{N_1},\Gamma_2=\bZ_{N_2}$. 
Then it is easy to verify that 
\begin{equation}
\eta_2=\eta_1^{-N_1/N_2}\left( \eta_1^{1/N_2}
-e^{-2\pi i k_2/N_2}\right)^{N_1}, 
\label{double1}
\end{equation}
with $k_2=0,1,\cdots,N_2-1$. 
This can be rewritten in the form 
\begin{equation}
\eta_1=\left( 1-e^{2\pi ik_1/N_1}\eta_2^{1/N_1}\right)^{-N_2}, 
\label{double2}
\end{equation}
with $k_1=0,1,\cdots,N_1-1$. 
It is important to notice that the form of these curves is 
unique for any $\alpha,\beta,\gamma,c_i$ that obey
(\ref{parameters}). 

\subsection{The gauge theory}

Consider now the low energy physics of M-theory on 
$X_{i,\Gamma_1\times\Gamma_2}$ (\ref{topology:d}).
Consider  
$X_{1,\Gamma_1\times\Gamma_2}=\bR^4/\Gamma_1\times \bS^3/\Gamma_2$.
The singularity 
of $\bR^4/\Gamma_1$ yields ${\cal N}=1$ super Yang-Mills 
theory with the gauge group $G_{\Gamma_1}$, 
where $G_{\Gamma_1}$ is the $ADE$ group that corresponds to $\Gamma_1$. 
The gauge bosons in four dimensions arise from integrating 
the $C$ field and wrapping M2-branes on the shrinking two-spheres
of $\bR^4/\Gamma_1$.
We need to interpret the low-energy physics of 
the second quotient $\bS^3/\Gamma_2$.
A natural interpretation is that this quotient corresponds to the
confining 
phase of ${\cal N}=1$ SYM with the gauge group $G_{\Gamma_2}$. 
As argued in \cite{amv}\cite{ach2}\cite{achvaf}, 
an M2-brane wrapping a non-trivial one-cycle in 
$\bS^3/\Gamma_2$ can be regarded as a QCD string of SYM. 
The domain wall that separates two distinct vacua is realized
by an  M5-brane wrapping the three-cycle $\bS^3/\Gamma_2$.  
Since we get a product of two  ${\cal N}=1$ super Yang-Mills 
theories in different phases, we 
can deduce that in this scenario 
the dynamical scales of the two theories obey 
$\Lambda_1\ll\Lambda_2$. 

The low-energy physics of M-theory on
$X_{2,\Gamma_1\times\Gamma_2}=\bR^4/\Gamma_2\times \bS^3/\Gamma_1$
is the same with interchanging
$G_{\Gamma_1}$ and  $G_{\Gamma_2}$.

Consider now 
$X_{3,\Gamma_1\times\Gamma_2}=\bR^4\times \bS^3/(\Gamma_1\times\Gamma_2)$.
This quotient space is not singular and admits no normalizable 
zero modes.
Thus, the low-energy physics of M-theory on
$X_{3,\Gamma_1\times\Gamma_2}$ 
is interpreted as the confining phase of 
$G_{\Gamma_1}\times G_{\Gamma_1}$ SYM theory. 
One support for this comes from the fact that, say for
$\Gamma_1=\bZ_{N_1},\Gamma_2=\bZ_{N_2}$ 
\begin{equation}
\pi_1(X_{3,\Gamma_1\times\Gamma_2})
=\pi_1(X_{3,\Gamma_1\times\Gamma_2}^{\prime})
=\bZ_{N_1}\times\bZ_{N_2} \ . 
\end{equation}
An M2-brane wrapping non-trivial 
one-cycles is a QCD string whose end point 
is a static quark that belongs to $(\N_1,\N_2)$ of 
$SU(N_1)\times SU(N_2)$.

The curve (\ref{double1}) and (\ref{double2}) 
exhibit  spontaneous breaking of the discrete quantum $R$-symmetry of SYM. 
For instance, consider $\eta_1\rightarrow e^{2\pi i}\eta_1$ in 
(\ref{double1}). 
A curve of label $k_2$ gets mapped to one 
with $k_2+1$. From (\ref{topology:d}), 
the phase of $\eta_1$ is identified with 
a vacuum angle of $SU(N_2)$ SYM. 
Thus, the transformation $\eta_1\rightarrow e^{2\pi i}\eta_1$ 
corresponds to a shift of the vacuum angle by $2\pi$. 
From field theory viewpoint, this shift amounts to 
a change of the phase factor of a gaugino bilinear condensate, 
in agreement with our interpretation of the curve. 
We see that for each $k_2=0,1,\cdots,N_2-1$, (\ref{double1}) 
describes a branch of the parameter space 
${\cal N}_{\Gamma_1\times\Gamma_2}$ that 
interpolates between the UV and
the confining phase of $SU(N_2)$ SYM with a gaugino 
bilinear condensate labelled by $k_2$. 
Similarly, the curve (\ref{double2}) exhibits 
spontaneous breaking of discrete $\bZ_{N_1}$ symmetry of $SU(N_1)$ 
SYM. 

Finally, note that since 
the different phases are smoothly interpolated, it
may be regarded as a support of the picture  
of low energy dynamics of SYM: mass gap, confinement 
and chiral symmetry breaking \cite{amv}\cite{atiwit}.

\section{Superpotentials and membrane instantons} 

In this section we will compute the $\cN=1$
superpotentials of M-theory on
the $G_2$ holonomy  quotients, and
comment on 
membrane instantons.

\subsection{Single quotient}

We start with the single quotient 
with $\Gamma_2=1$. 
The M-theory curve reads \cite{atiwit}
\begin{equation}
\eta_2=\eta_1^{-N_1}\prod_{t_1,\mu_1}
\Big(\eta_1-\exp(2\pi i\mu_1/t_1)\Big)^{t_1h_{t_1}},~~
\eta_3=\prod_{t_1,\mu_1}
\Big(1-\exp (-2\pi i\mu_1/t_1)\eta_1\Big)^{-t_1h_{t_1}} \ , 
\label{curve;single}
\end{equation}
and the structure of zeros and poles is given by:
\begin{equation}
\begin{array}{|c|c|c|c|} \hline 
 & P_1^{t_1,\mu_1} & P_2 & P_3 \\ \hline
\eta_1 & e^{2\pi i\mu_1/t_1} & \infty & 0 \\ \hline
\eta_2 & 0^{t_1h_{t_1}} & 1 & \infty^{N_1} \\ \hline
\eta_3 & \infty^{t_1h_{t_1}} & 0^{N_1} & 1 \\ \hline
\end{array}
\end{equation}

$\eta_i$ are meromorphic 
functions on a sphere. 
Let $z$ be the coordinate of this sphere. 
We map the points $P_2,P_3$ to
$z=\omega,\omega^2$ respectively, 
with $\omega=e^{2\pi i/3}$. 
$\eta_1$ can be written in the form
\begin{equation}
\eta_1=-\omega{z-\omega^2 \over z-\omega} \ . 
\label{parameterrep1}
\end{equation}
In this parametrization
the point $P_1^{t_1,\mu_1}$
corresponds to 
\begin{equation}
z=\alpha_{t_1,\mu_1}=\omega\, 
{e^{2\pi i\mu_1/t_1}+\omega^2\over e^{2\pi i\mu_1/t_1}+\omega}
={\cos \left({\pi\mu_1\over t_1}+{\pi\over 3} \right) \over 
  \cos \left({\pi\mu_1\over t_1}-{\pi\over 3} \right)} \ . 
\end{equation}

It follows from (\ref{curve;single}) that 
\begin{eqnarray}
&&\eta_2=c^{\prime}\,{
\prod_{t_1,\mu_1}(z-\alpha_{t_1,\mu_1})^{t_1h_{t_1}}
           \over (z-\omega^2)^{N_1}}, \nn
&&\eta_3=c^{\prime\prime}\,{(z-\omega)^{N_1}\over 
\prod_{t_1,\mu_1}(z-\alpha_{t_1,\mu_1})^{t_1h_{t_1}}} \ , 
\label{parameterrep2}
\end{eqnarray}
where 
\begin{equation}
c^{\prime}=
\prod_{t_1,\mu_1}(1+\omega^{-1}e^{2\pi i\mu_1/t_1})^{t_1h_{t_1}} , ~~
c^{\prime\prime}=
\prod_{t_1,\mu_1}(\omega + e^{2\pi i\mu_1/t_1})^{-t_1h_{t_1}} \ . 
\end{equation}
Note that the single quotient preserves the subgroup
$\bZ_2^{(1)}\subset \Sigma_3$ that acts on $\eta_i$ as 
\begin{equation}
\bZ_2^{(1)}:~~(\eta_1,\eta_2,\eta_3)\rightarrow 
(\eta_1^{-1},\eta_3^{-1},\eta_2^{-1}) \ ,
\end{equation}
which is equal to the action $z\rightarrow 1/z$.

Consider now the
superpotential. 
The quantum corrections to 
the superpotential come from instantons that have two fermionic zero 
modes. They correspond to euclidean membranes 
wrapping associative three-cycles. 
These contributions should vanish at $P_i$ since
$P_i$ correspond to the large volume (classical gauge theory) limit. 
Thus, we conclude that $W$ is proportional to
\begin{equation}
f(z)\equiv (z-\omega)(z-\omega^2)\prod_{t_1,\mu_1}
(z-\alpha_{t_1,\mu_1}) \ .
\end{equation}
We assume that $W$ has no other zeros.
Consider, for instance,  the behavior near the point $P_1$.
We write
\begin{equation}
z\sim \alpha_{t_1,\mu_1}+\tilde{u} \ , 
\end{equation}
with $\tilde{u}= u^{1/t_1h_{t_1}}$ being a good local coordinate around 
$P_1^{t_1,\mu_1}$. 

The superpotential $W$ behaves as 
\begin{equation}
W\sim \tilde{u} \ , 
\end{equation}
which is consistent with the fact that
$\tilde{u}$ has been identified 
with a gaugino bilinear condensate of SYM 
with gauge group $K_{t_1}$. 
So the above result is in accord with a field theoretic result. 

Since $W(z)$ is a meromorphic function on the sphere, 
$W(z)$ has the same order of zeros and poles.
Thus, it takes the form 
\begin{equation}
W(z)=ic\,{f(z) \over g(z)} \ , 
\end{equation}
where $g(z)$ is a polynomial of degree $l\equiv 2+\sum_{t_1,\mu_1}1$, 
and $c$ is a constant. 

In order to determine $g(z)$, we recall that the single quotient 
preserves the discrete symmetry $\bZ_2^{(1)}\subset \Sigma_3$ 
that acts on the superpotential as $R$-symmetry 
\begin{equation}
W(1/z)=-W(z) \ . 
\end{equation}
Using the relation
\be
\prod_{\mu_1}-\alpha_{t_1,\mu_1}=1~~{\rm for}~t_1=2,3,4,5 \ , 
\ee
we find that
\begin{equation}
f(1/z)=-z^{-l}f(z) \ .
\end{equation}
Thus, $g$ obeys
\begin{equation}
g(1/z)=z^{-l}g(z) \ . 
\end{equation}

When $\Gamma_1=\bZ_{N_1}$, $W$ takes the form 
\begin{equation}
W(z)=ic\,{z^3-1 \over 
z^3+\xi z^2+\xi z+1} \ . 
\end{equation}
Here $\xi$ is a constant and $\xi(N_1=1)=0$ in order to be 
consistent with \cite{atiwit}. 
We can rewrite the superpotential in terms of 
the couplings $\eta_i$, with manifest $\bZ_2^{(1)}$ symmetry 
\begin{equation}
W=-ic(\omega-\omega^{-1})\,
{\eta_1+\eta_2^{1/N_1}+\eta_3^{1/N_1}+
 \eta_1^{-1}+\eta_2^{-1/N_1}+\eta_3^{-1/N_1} 
\over 
(1-2\xi) \left(\eta_1-\eta_1^{-1}\right)
+(1+\xi) \left(\eta_2^{1/N_1}+\eta_3^{1/N_1}-
               \eta_2^{-1/N_1}-\eta_3^{-1/N_1}\right)} \ .
\end{equation}

Let us study now the membrane instanton corrections to $\eta_i$,
where we  
consider the case $\Gamma_1=\bZ_{N_1}$. 
We have 
\begin{equation}
{z^3-1 \over 
z^3+\xi z^2+\xi z+1}=u \ , 
\end{equation}
with $u$ an instanton factor. 

Let $z_1(u),z_2(u),z_3(u)$ be the solutions to this equation 
such that $z_i(u=0)=\omega^{i+2}$. 
$u$ in $z_1$ 
comes from a fractional M2 instanton of instanton number $1/N_1$. 
On the other hand, $u$ in $z_2,z_3$ is due to an ordinary M2 
instanton wrapping the associative three cycle $\bS^3/\bZ_{N_1}$. 
By plugging the solutions into (\ref{parameterrep1}) and 
(\ref{parameterrep2}), the coupling constants $\eta_i$ can be 
written in terms of an power expansion in $u$. 
The instanton contributions to $\eta_1$ around $P_1$
take the form 
\begin{eqnarray}
\eta_1(z=z_1(u))=1-{2i(\xi+1) \over 3\sqrt{3}}\, u
-{2(\xi+1)^2 \over 27}\,u^2+\cdots.
\end{eqnarray}
M2 instanton contributions 
to $\eta_2$ take the form
\begin{equation}
\eta_2(z=z_2(u))=1+{N_1i(\xi-2)\over 3\sqrt{3}}\,u-
{N_1\Big((\xi-2)N_1-3\xi\Big)(\xi-2)\over 54}\,u^2+\cdots. 
\end{equation}

\subsection{Double quotient}

Consider the double quotient and let
$\Gamma_1=\bZ_{N_1},\Gamma_2=\bZ_{N_2}$. 
We have
\begin{equation}
\eta_1=c_1\,{(z-\gamma)^{N_2}\over (z-\beta)^{N_2}}, ~~
\eta_2=c_2\,{(z-\alpha)^{N_1}
             \over (z-\gamma)^{N_1}}, ~~
\eta_3=c_3\,{(z-\beta)^{N_1N_2}\over (z-\alpha)^{N_1N_2}} \ . 
\end{equation}
For simplicity, we take 
\begin{equation}
\alpha=1,~~\beta=\omega,~~\gamma=\omega^2 \ . 
\end{equation}
As in the case of the single quotient, the superpotential must 
vanish at $P_1,P_2,P_3$. 
It is a meromorphic function on the sphere and therefore the order
of its poles is three.
Thus, it
takes the form 
\begin{equation}
W=ic\,{z^3-1\over z^3+\xi_1 z^2+\xi_2 z+\xi_3} \ . 
\end{equation}
Note that
the double quotient breaks 
$\bZ_2^{(1)}$. 
As a check, consider the behavior around $P_1$ by setting 
\begin{equation}
z\sim 1+u \ . 
\end{equation}
The superpotential is given by 
\begin{equation}
W\sim u \ . 
\end{equation}
Since $u\sim \Lambda_1^3$, this can be identified with 
the superpotential of $SU(N_1)$ SYM. 
Indeed, it is consistent that no dependence of $\Lambda_2^3$ 
appears, since
we are working in the regime $\Lambda_1\gg\Lambda_2$.

\section{Double quotients with Wilson lines}

In this section we will make some comments and present some questions
on Wilson lines and the M-theory moduli space. 
We start with M-theory on the singular $G_2$ holonomy
manifold 
$X_{1,\Gamma_1\times\Gamma_2}=\bR^4/\bZ_{N_1}\times \bS^3/\bZ_{N_2}$. 
The low energy dynamics is described by a seven-dimensional $SU(N_1)$ 
SYM compactified on $Q=\bS^3/\bZ_{N_2}$. 
As usual, supersymmetry is preserved by a twist, 
namely the identification of spin connections 
$w_{SO(3)_R}=w_{SO(3)_Q}$. The bosonic matter content of the 
resulting theory is found to be 
\begin{eqnarray}
\begin{array}{ccc}
     &SO(1,3)&SO(3) \\
A    &{\bf 4} &{\bf 1}   \\
\phi &{\bf 1} &{\bf 3}
\end{array}
\end{eqnarray}
$\phi$ is a complex scalar which belongs to the adjoint 
representation of $SU(N_1)$. 
Now we turn on Wilson lines for $\phi$ along a one-cycle $\gamma$ 
of $\bS^3/\bZ_{N_2}$ 
\begin{equation}
g_{\gamma}={\rm P}\exp i\oint_{\gamma}\phi
={\rm diag}(I_{n_1},\,e^{2\pi i/N_2}I_{n_2},\,\cdots,
                     \,e^{2\pi i(k-1)/N_2}I_{n_k}),\quad k\le N_2 \ , 
\end{equation}
with $\sum_in_i=N_1$ (see also \cite{wit:wl}\cite{gopvaf:wl}).
$g_\gamma$ is an $N_1$-dimensional reducible representation 
of $\bZ_{N_2}$. 
Recall that
the Cartan part of $\phi$ is given by 
\begin{equation}
\phi^a=\int_{\beta_a} C,\quad a=1,2,\cdots, N_1-1 \ . 
\end{equation}
Here $\beta_a$ are collapsing two-cycles in $\bR^4/\bZ_{N_1}$. 
Thus,
the Wilson lines are given by integrating the
$C$-field on three-cycles $C_a=\beta_a\times\gamma$. 

The Wilson line breaks the gauge group $SU(N_1)$ to 
$\prod_i SU(n_i)$.  
The mass of the W-bosons is of order $r_0^{-1}$, where $r_0$
is the size of the one-cycle in 
$\bS^3/\bZ_{N_2}$. 
Alternatively, the seven-dimensional SYM 
has the interaction term 
\begin{equation}
\int d^7x\, {\rm tr}\!\left[ \phi, A \right]^2 \ , 
\end{equation}
which gives mass to $A$.
Since we are working in the low energy regime
$E \ll r_0^{-1}$, the massive modes may be neglected and
the gauge theories decouple from one another. 
In the context of M-theory, this Higgs mechanism can be interpreted 
as follows.
The gauge bosons of $SU(N_1)$ come from 
M2-branes wrapping collapsing cycles $\beta$ in $\bR^4/\bZ_{N_1}$. 
The Wilson lines turn on a $C$ field, 
which gives mass to some of the wrapped membranes. 
Consequently we have $\prod_i SU(n_i)$ symmetry rather 
than $SU(N_1)$. 

Consider now $X_{2,\Gamma_1\times\Gamma_2}$ and
$X_{3,\Gamma_1\times\Gamma_2}$
without turning on Wilson lines.
As we have seen, these two geometries can be interpolated
smoothly.
It is unlikely that
we interpolate between them and $X_{1,\Gamma_1\times\Gamma_2}$
with Wilson lines.
A natural guess is that with a Wilson line 
we get a new branch of 
the moduli space that is disconnected with the branch without 
Wilson lines, and we view the Wilson lines as discrete moduli. 
If such a branch exists, it is interesting to know
whether there is a $G_2$ flop that 
interpolates $X_{1,\Gamma_1\times\Gamma_2}$ with Wilson lines 
smoothly to a $G_2$ manifold that 
describes the vacuum structure of $\cN=1$ $\prod SU(n_i)$ SYM. 
For the latter, 
the geometry should possess no $ADE$ singularity 
because of a mass gap of SYM. 
It should also 
account for the Witten index 
of $\cN=1$ $\prod SU(n_i)$ SYM 
together with that of ${\cal N}=1~SU(N_2)$ SYM 
\begin{equation}
{\rm tr}\,(-1)^F=N_2 \prod n_i \ . 
\label{w;index}
\end{equation}

A candidate geometry is 
$X=\bR^4\times\{\bS^3/(\bZ_{n_1}\times\bZ_{N_2})\#\cdots\#\
\bS^3/(\bZ_{n_k}\times\bZ_{N_2})\}
$, 
namely a ``multi-center'' $G_2$ holonomy manifold such that 
around each center it looks like 
$\bR^4\times\bS^3/(\bZ_{n_i}\times\bZ_{N_2})$. 
$\bZ_{N_2}$ acts diagonally. 
The number of discrete vacua can be seen by looking at 
a fractional change of $C$-field fluxes on associative three-cycles 
under a shift of the vacuum angle 
$\theta$ of SYM. 
Indeed, one finds that 
\begin{equation}
\int_{\bZ_{n_i}\!\setminus\,\bS^3/\bZ_{N_2}}C\rightarrow 
\int_{\bZ_{n_i}\!\setminus\,\bS^3/\bZ_{N_2}}C\,+{2\pi \over n_i} \ .
\end{equation}
under the shift of the vacuum angle of $\cN=1~\prod SU(n_i)$ SYM 
by $2\pi$. 
We also find that 
$\pi_1(X)=\bZ_{N_2}\times\bZ_{n_1}\times\bZ_{n_2}\times\cdots$. 
This is consistent with the interpretation that 
an M$2$-brane wrapping non-trivial one-cycles of $X$ is a QCD string 
whose end point belongs to $({\bf N_2},{\bf n_1},{\bf n_2},\cdots)$
of $SU(N_2)\times \prod SU(n_i)$. 
Note also that KK reduction of this geometry along the Hopf fibers of 
$\bZ_{n_i}\!\!\setminus\!\bS^3/\bZ_{N_2}$, 
yields a generalized conifold \cite{gubneksha}
that takes the form of multi small resolved conifolds 
with RR-fluxes on the Hopf bases $\bS^2/\bZ_{N_2}$. 
The mirror of this CY is reminiscent of a setup 
studied in \cite{cacintvaf} 
in the context of
SYM with a product group via 
a geometric transition.

\section{Discussion}

In the paper we extended results of \cite{atiwit}
to more general quotients of $S(\bS^3)$, and in particular 
the ``double'' quotient, 
where the quotient group is a subgroup of the isometry of 
$S(\bS^3)$. We analysed 
the moduli space of M-theory on the quotient and show
that the moduli space interpolates smoothly between three different 
types of classical geometries.
These three geometries correspond to different low energy dynamics. 
We computed the exact superpotentials of M-theory on the
quotients of 
$S(\bS^3)$, and  
discussed the
membrane instanton physics. 
We commented 
on turning on 
Wilson lines and breaking of the gauge symmetry 
and discussed some of the implications.

There are several issues that deserve further study.
One is to derive equation (\ref{anomaly}) 
by an anomaly analysis. 
Another issue is to understand in the context of M-theory
dynamics the 
singularities of the superpotentials computed in section 4.
It would also be interesting to know the precise relation 
between the complexified gauge coupling of SYM $\tau$ and
$\eta_i$, 
$\tau=\tau(\eta_i)$. 
Once given, the analysis of section 4 will provide us 
with information about nonperturbative corrections
to the $\cN=1$ SYM coupling. 

Finally, the structure of
the moduli space of M-theory on the $G_2$ holonomy manifolds
with Wilson lines is still largely unclear and deserves further study. 
From the discussion of the 
previous section, one may deduce a duality of 
superstrings, which is an extension of \cite{vafa}. 
The KK reduction of 
$\bR^4/\bZ_{N_1}\times \bS^3/\bZ_{N_2}$ along an 
$\bS^1$ in the 
ALE fibration leads to $N_1$ D6 branes wrapping $\bS^3/\bZ_{N_2}$ 
of a deformed conifold. 
On the other hand, 
from KK reduction of 
$X=\bR^4\times\{\bS^3/(\bZ_{n_1}\times\bZ_{N_2})\#\cdots\#\
\bS^3/(\bZ_{n_k}\times\bZ_{N_2})\}
$, 
along the Hopf fibers of 
$\bZ_{n_i}\!\!\setminus\!\bS^3/\bZ_{N_2}$,
one obtains superstring theory on the resultant CY.
We may conjecture that both are dual to each other. 
One can
try to verify the duality in the context of topological strings 
\cite{gopvaf:dual}. 
The superstring theory with the D$6$ 
branes wrapping the three-cycle of a deformed conifold 
reduces to 
a topological open string that is
given by $SU(N_1)$ Chern-Simons theory on $\bS^3/\bZ_{N_2}$ with 
Wilson lines. 
It would be nice to see if this theory is dual to a topological 
closed string on the CY: 
\begin{center}
\begin{tabular}{ccc}
$\bR^4/\bZ_{N_1}\times \bS^3/\bZ_{N_2}$ & 
$\stackrel{\rm G_2\,flop}{\Longleftrightarrow}$ & 
$X$ \\
{\footnotesize with Wilson lines}              & &    \\
$\Downarrow$ &  & $\Downarrow$ \\
$N_1$ D$6$ branes & 
$\stackrel{\rm dual}{\Longleftrightarrow}$ &
superstring on CY \\
{\footnotesize wrapping a three-cycle of a deformed conifold}  & & \\
{\footnotesize with Wilson lines} && \\
$\Downarrow$ &  & $\Downarrow$ \\
topological open string & $\stackrel{\rm dual}{\Longleftrightarrow}$ & 
topological closed string on CY \\
$SU(N_1)$ CS on $\bS^3/\bZ_{N_2}$ & & \\
{\footnotesize with Wilson lines}              & &    
\end{tabular}
\end{center}

\vskip 1cm

\section*{Acknowledgements}

This research is supported by the US-Israel Binational Science
Foundation.

\newpage


\end{document}